\documentstyle[sprocl]{article}
\input{psfig}
\def\be{\begin{equation}}
\def\ee{\end{equation}}
\def\bea{\begin{eqnarray}}
\def\eea{\end{eqnarray}}

\def\beq{\begin{equation}}
\def\eeq{\end{equation}}
\def\eq{\end{equation}}
\def\to{\rightarrow}

\def\sigv{\langle \sigma v\rangle}

\def\bsg{\ifmmode B\to X_s\gamma\else $B\to X_s\gamma$\fi}
\def\bsll{\ifmmode B\to X_s\ell^+\ell^-\else $B\to X_s\ell^+\ell^-$\fi}
\def\bstt{\ifmmode B\to X_s\tau^+\tau^-\else $B\to X_s\tau^+\tau^-$\fi}
\def\shat{\ifmmode \hat{s}\else $\hat{s}$\fi}

\newcommand{\newc}{\newcommand}

\newc{\lcal}{\int {\cal L}dt}

\newc{\mHpm}{m_{H^\pm}}
\newc{\gsim}{\lower.7ex\hbox{$\;\stackrel{\textstyle>}{\sim}\;$}}
\newc{\lsim}{\lower.7ex\hbox{$\;\stackrel{\textstyle<}{\sim}\;$}}
\newc{\ie}{{\it i.e.}}          
\newc{\etal}{{\it et al.}}
\newc{\eg}{{\it e.g.}}          
\newc{\kev}{\hbox{\rm\,keV}}            
\newc{\mev}{\hbox{\rm\,MeV}}            
\newc{\gev}{\hbox{\rm\,GeV}}            
\newc{\tev}{\hbox{\rm\,TeV}}
\newc{\xpb}{\hbox{\rm\, pb}}
\newc{\xfb}{\hbox{\rm\, fb}}

\newc{\mtop}{m_t}
\newc{\mbot}{m_b}
\newc{\mz}{m_Z}
\newc{\mw}{M_W}
\newc{\alphasmz}{\alpha_s(m_Z^2)}
\newc{\swsq}{\sin^2\theta_W}
\newc{\tw}{\tan\theta_W}
\newc{\cw}{\cos\theta_W}
\newc{\sw}{\sin\theta_W}
\newc{\BR}{\hbox{\rm BR}}
\newc{\zbb}{Z\to b\bar}
\newc{\Gb}{\Gamma (Z\to b\bar b)}
\newc{\Gh}{\Gamma (Z\to \hbox{\rm hadrons})}
\newc{\rbsm}{R_b^\hbox{\rm sm}}
\newc{\rbsusy}{R_b^\hbox{\rm susy}}
\newc{\drb}{\delta R_b}

\newc{\sgn}{\mbox{sgn}}
\newc{\tbeta}{\tan\beta}
\newc{\uL}{{\tilde u_L}}
\newc{\uR}{{\tilde u_R}}
\newc{\cL}{{\tilde c_L}}
\newc{\cR}{{\tilde c_R}}
\newc{\tL}{{\tilde t_L}}
\newc{\tR}{{\tilde t_R}}
\newc{\dL}{{\tilde d_L}}
\newc{\dR}{{\tilde d_R}}
\newc{\sL}{{\tilde s_L}}
\newc{\sR}{{\tilde s_R}}
\newc{\bL}{{\tilde b_L}}
\newc{\bR}{{\tilde b_R}}
\newc{\eL}{{\tilde e_L}}
\newc{\eR}{{\tilde e_R}}
\newc{\mhp}{m_{H^\pm}}
\newc{\mhalf}{m_{1/2}}

\newc{\lR}{\tilde{l}_R}
\newc{\lL}{\tilde{l}_L}
\newc{\nL}{\tilde{\nu}_L}
\newc{\na}{\chi^0_1}
\newc{\nb}{\chi^0_2}
\newc{\nc}{\chi^0_3}
\newc{\nd}{\chi^0_4}
\newc{\ca}{\chi^{\pm}_1}
\newc{\cb}{\chi^{\pm}_2}
\newc{\camp}{\chi^\mp_1}
\newc{\cbmp}{\chi^\mp_1}
\newc{\capos}{\chi^{+}_1}
\newc{\caneg}{\chi^{-}_1}
\newc{\phit}{\phi_t}
\newc{\phib}{\phi_b}
\newc{\phiew}{\phi_{ew}}
\newc{\htz}{h^0_t}
\newc{\hbz}{h^0_b}
\newc{\hewz}{h^0_{ew}}
\newc{\hsmz}{h^0_{sm}}

\def\NPB#1#2#3{Nucl. Phys. B {\bf #1}, #3 (19#2)}
\def\PLB#1#2#3{Phys. Lett. B {\bf #1}, #3 (19#2)}

\def\PRD#1#2#3{Phys. Rev. D {\bf #1}, #3 (19#2)}
\def\PRL#1#2#3{Phys. Rev. Lett. {\bf#1}, #3 (19#2)}

\def\beq{\begin{equation}}
\def\eeq{\end{equation}}
\def\slashchar#1{\setbox0=\hbox{$#1$}           
   \dimen0=\wd0                                 
   \setbox1=\hbox{/} \dimen1=\wd1               
   \ifdim\dimen0>\dimen1                        
      \rlap{\hbox to \dimen0{\hfil/\hfil}}      
      #1                                        
   \else                                        
      \rlap{\hbox to \dimen1{\hfil$#1$\hfil}}   
      /                                         
   \fi}                                         %
\catcode`@=11
\long\def\@caption#1[#2]#3{\par\addcontentsline{\csname
  ext@#1\endcsname}{#1}{\protect\numberline{\csname
  the#1\endcsname}{\ignorespaces #2}}\begingroup
    \small
    \@parboxrestore
    \@makecaption{\csname fnum@#1\endcsname}{\ignorespaces #3}\par
  \endgroup}
\catcode`@=12

\begin{document}
\title{MASS DENSITY OF NEUTRALINO DARK MATTER}
\author{ James D.\ Wells\footnote{Work supported by DOE
under Contract DE-AC03-76SF00515. SLAC-PUB-7606. }}
\address{Stanford Linear Accelerator Center, Stanford University \\
Stanford, California 94309, USA }
\maketitle\abstracts{
The lightest supersymmetric particle (LSP) is stable in an $R$-parity
conserving theory.  In this article the steps needed to calculate
the present day mass density of such a particle are detailed.
It is shown that there can be a significant amount of LSP dark
matter in the universe.  Furthermore, relic abundance considerations
put an upper bound on how large supersymmetry breaking masses
can be without resorting to finetuning arguments.}


\section{Introduction}

The most general gauge invariant superpotential will generally lead
to unacceptable fast proton decay.  For this reason, a discrete symmetry
must be posited that banishes baryon and/or lepton violating operators
which contribute to this decay.  The simplest such discrete symmetry
is $R$-parity~\cite{rparity}.  Exact $R$-parity conservation makes the lightest
supersymmetric partner (LSP) stable, thereby introducing many interesting
phenomena.  The most studied phenomenon is the large missing energy
signature associated with production of superpartners and subsequent
decay into the LSP 
plus jets, photons, leptons, etc.  For experimental
reasons~\cite{falk94:248} and theoretical reasons~\cite{diehl95:4223}
the LSP is now expected to be the lightest neutralino.

A stable LSP has more than collider physics consequences.  If they are
created in the hot and violent early days of the universe, then there
should be some left over today, and perhaps they could have significant
cosmological and astrophysical
consequences.  If it turns out that the LSP constitutes
a significant mass fraction of the universe then it should be possible
to witness large-scale gravitational effects of these particles on
galaxies and clusters.  Experiments have been testing large-scale
gravity for decades now, and the current consensus states that there
are non-luminous sources of gravitational import beyond the ordinary
baryonic matter that makes up planets and stars~\cite{trimble87:425}.  
Part of the evidence
of additional mass-energy beyond our luminous matter
includes rotation curves of galaxies and infall of clusters.  The evidence
for non-baryonic dark matter can be attributed to the successful
agreement between measurement and theory in big bang nucleosynthesis (BBN).
BBN tells us that baryonic mass fraction of the universe is probably less
than 10\%~\cite{schramm}.  (I am making the usual assumption 
that the total energy
density of the universe is the critical energy density in accord with
a general inflation scenario.)

One is left wondering what the rest of the universe is made of.  One
suggestion is a weakly interacting massive object, or WIMP.  The
acronym WIMP is a good general label, but it is a bit misleading.
Upon closer inspection a particle of 
${\cal O}(m_W)$ mass which has full-strength $SU(2)$ (weak) interactions
is generally not a good dark matter candidate, and yields a relic
abundance less than is required to have astrophysical significance.
Some additional suppressions are generally needed in the annihilation
cross-section.  As we will see in subsequent sections, the LSP generally
has the right suppressions to make it a good dark matter candidate
while at the same time having mass near the weak-scale.  For this reason,
I will not use the generic word WIMP, and instead refer to the dark
matter candidate as the LSP.  
LSPs are
excellent candidates for the dark matter because 
{\it if enough can be around} they allow conformance
to experimentally determined properties of galactic rotation
curves, structure formation and
big bang nucleosynthesis.  Furthermore, calculation of their relic
abundance indicates that LSPs could contribute most of the energy
density of the universe.  This is a non-trivial separate test that
confirms that {\it enough can be around} to solve the observational
problems.

In the lightest neutralino, 
supersymmetry provides a natural dark matter candidate.
In fits of optimism one could even declare the galactic rotation
curve anomalies, etc.~as positive experimental evidences for
supersymmetry.  However, in this chapter the focus will be
on two main topics.  First, and foremost, I will provide the
details on how to calculate the relic abundance of a weakly
interacting particle.  Since obtaining the correct relic abundance
is a somewhat involved calculation, but also an important
one, it is useful to have detailed discussion.
Second, I will tailor some additional
remarks about the calculation to the lightest neutralino, and
show how the results constrain {\it other}
supersymmetric particle masses.  In general it can be shown
that there is an upper limit to superpartner masses due
to relic abundance considerations alone.  The limit is
based on physical necessity 
and not on arbitrary fine-tuning considerations.  This insight
is as important as the realization that supersymmetry can
cure the ``dark matter problem.''  

The subsequent sections reflect the goals presented in the previous
paragraph.  Much of the emphasis will be placed in the techniques 
of calculating the relic abundance.  I have attempted to include in
this one source all the necessary general relativity, statistical
mechanics, and particle physics knowledge needed to follow
a precise calculation.  I have also included a section which
derives an accurate and approximate solution to the Boltzmann
equation.  These results will then be used to analyze quantitatively
how the supersymmetric spectrum is affected by the relic abundance
constraint.

\section{Solving the Boltzmann equation}

The starting point is the Boltzmann equation,
\beq
\frac{dn}{dt}= -3 H n-\langle \sigma v\rangle (n^2-n^2_{eq})
\eeq
where $H$ is the Hubble constant, $n$ is the particle number
density in question, $n_{eq}$ is the particle number equilibrium density,
and $\langle \sigma v\rangle$ is the thermal averaged cross-section.
The Boltzmann equation is simple in form but somewhat subtle to solve.
The differentiating parameter is time $t$, however $n$ and $n_{eq}$ are
most easily characterized by temperature.  Furthermore, the Hubble constant
evolution is best traced by the relative time change in the scale
parameter $H=\dot a/a$.  Only one parameter is independent, and so 
the first step will be to cast the 
Boltzmann equation into a purely temperature dependent 
relation.  The two equations
that will allow us to do that are the ``Friedmann equation'' and the
``conservation of entropy equation.''


The use of Einstein's equation of general relativity is necessary
to reveal the explicit time dependence of $\dot a/a$.  This familiar
equation states that
\beq
R_{\mu\nu}-\frac{1}{2}Rg_{\mu\nu}=8\pi G_N T_{\mu\nu}.
\eeq
To do explicit calculations the metric and stress-energy must
be defined.  We use the flat Robertson-Walker metric
\beq
ds^2=dt^2-a^2(t)\left[ dx^2+dy^2+dz^2\right]
\eeq
which assumes the universe to be flat, homogeneous, and isotropic.
This is equivalent to the metric tensor 
\beq
g_{\mu\nu}=\mbox{diag}(1,-a^2(t),-a^2(t),-a^2(t)).
\eeq
The general stress-energy consistent with an homogeneous and isotropic
universe is
\beq
T_{\mu\nu}={\mbox{diag}}(\rho,-p,-p,-p).
\eeq
Einstein's equation is of course valid for each $\{\mu\nu\}$, however we only
need the $00$ component.  From the metric tensor it is 
straightforward~\cite{kolb90:1} to
compute the Ricci tensor component $R_{00}$ and the Ricci scalar $R$: 
\bea
R_{00} & = & -3\frac{\ddot a}{a} \\
R & = & -6\left( \frac{\ddot a}{a}+\frac{\dot a^2}{a^2}\right).
\eea
(Dots indicate time derivative.)
The $00$ component of the Einstein equation under the flat Robertson-Walker
metric is then simply,
\beq
\frac{\dot a}{a} =\frac{\sqrt{\rho}}{\kappa}~~\mbox{where}~~
\kappa = \sqrt{\frac{3}{8\pi G_N}} .
\eeq
This equation is often called the Friedmann equation.  This will be
quite useful to get rid of the scale factor in the Boltzmann equation.
The value of $\rho$ on the right hand side of the equation will be 
calculated later and is conveniently parameterized by its temperature
dependence.  Thus, the Friedmann
equation provides a nice connection between the time dependent 
relative scale factor (the Hubble constant) 
and the temperature.



In thermal equilibrium entropy is conserved, and using
the first law of thermodynamics we can
identify the entropy as $S(T)=(\rho +p)V/T$ up to an irrelevant
constant.
With $V=a^3$ we define the entropy density $s(T)$ as
\beq
s(T)=\frac{S(T)}{a^3}=\frac{\rho +p}{T}.
\eeq
The conservation of entropy means that $S(T)=s(T)a^3$ is time independent:
\beq
\frac{d}{dt}(s(T)a^3)=0 ~~\Longrightarrow ~~
\dot T = -3\frac{\dot a}{a} \frac{s(T)}{s'(T)}.
\eeq
(The prime on $s'(T)$ indicates a temperature derivative.)
This equation is a direct result of the conservation of entropy in
thermal equilibrium and so is called the ``conservation of entropy
equation''.  Its utility is relating the time derivative of
the temperature ($\dot T$) to the scale factor time derivative 
($\dot a/a$).



We actually have enough information from the above paragraphs
to construct the temperature dependent Boltzmann equation.  
First, we rewrite $dn/dt=(dn/dT)\dot T$ and use the ``conservation of
entropy'' equation to replace $\dot T$ in favor of $\dot a/a$.
Then we use the ``Friedmann equation'' to replace
$\dot a/a$ in favor of the energy density $\rho (T)$.  The result
is
\beq
\label{bolt2}
\frac{dn}{dT}=\frac{s'(T)}{s(T)}\left\{ n+\frac{\kappa J(t)}{3\sqrt{\rho(T)}}
 [n^2-n^2_{eq}(T)]\right\}
\eeq
where $J(T)=\langle \sigma v\rangle (T)$ is the thermal averaged
cross-section.

This equation might not look like much progress; however, all non-trivial
dependences from $\rho (T)$ and $s(T)$ are easily calculated as  
functions of temperature.  This will be demonstrated below.
At sufficiently high temperature ($T_H$),
where all relevant particles are in
thermal equilibrium, then we know as a boundary condition that
$n(T_H)=n_{eq}(T_H)$.  One then need only integrate down to today's
temperature ($T\simeq 0$) to obtain the current number density in
the universe $n(0)$.  The mass density is then just $\rho_\chi =m_\chi n(0)$
where $m_\chi$ is the mass of the relic particle of interest.

Often it is of interest to compare the mass density of our
relic particle to the critical density $\rho_c=\kappa^2 H^2$ needed for
a flat universe.  The relevant formula is
\beq
\Omega_\chi = \frac{m_\chi n(0)}{\kappa^2 H^2}.
\eeq
For a flat universe the sum of all contributing $\Omega$'s 
(baryons, neutrinos, cosmological constant, neutralinos, etc.) must
be equal to 1. If a massive stable particle makes up a significant 
fraction of the total critical density then it is an interesting
cold dark matter candidate.  If the calculated critical density is
too high, then it is said to
``overclose the universe'', meaning that the universe became matter
dominated too early and it is impossible to reconcile the current
mass density and the Hubble constant with the age of the universe.



In order to effectively solve the Boltzmann equation we must have
an understanding of the thermodynamic quantities which enter the
equation.  These are the equilibrium number density $n_{eq}(T)$,
the energy density $\rho(T)$, and the
entropy density $s(T)$.  Since $s(T)=(\rho +p)/T$ we can focus on
calculating the pressure $p(T)$ rather than computing $s(T)$ directly.
Calculating thermodynamic quantities is standard
statistical mechanics and can be found in numerous sources.
Here I will merely argue the most salient points that will lead
to workable equations. 

The density of states in a phase space volume $d^3\vec xd^3\vec k$
is $1/(2\pi)^3$.  Integrating over the volume, the phase space volume
density of states is
\beq
d\xi = \frac{V}{(2\pi)^3}d^3\vec k =\frac{V}{(2\pi )^3} (4\pi) k^2dk,
\eeq
where $k=|\vec k |$ here.
We are interested to begin with 
in the number of particles per unit volume (number density).  It will
be necessary to multiply the density of states times the mean occupation
number for a given momentum state $|\vec k\rangle$.  The mean occupation
number is expressed by the Fermi and Bose distribution functions,
\beq
f_\eta(k,T) =\frac{1}{e^{\sqrt{k^2+m^2}/T}+\eta}
\eeq
where
$\eta =-1,1$ for bosons and fermions respectively.  Thus, the total
number density integrated over all momentum modes is
\beq
n_{eq}(T)=\frac{g}{V}\int d\xi f_\eta (k,T) 
 =\frac{g}{2\pi^2}\int_0^{\infty} dk k^2 f_\eta(k,T)
\eeq
where $g$ is the number of internal spin degrees of freedom.  

Similarly, we can calculate the energy density and pressure as
\bea
\rho(T) & = & \frac{g}{2\pi^2}\int_0^\infty dk k^2 \sqrt{k^2+m^2}
     f_\eta(k,T) \\
p(T) &=& \frac{g}{2\pi^2}\int_0^\infty dk \frac{k^4}{\sqrt{k^2+m^2}}
 f_\eta(k,T).
\eea
Since $s(T)=(\rho(T)+p(T))/T$ we are done.  We should keep in mind that
$n_{eq}(T)$ in the Boltzmann equation 
only applies for the one relic particle.  However, the $\rho(T)$ and
$s(T)$ in the equation are the total energy density and entropy density
summed over all particles.

We always want $T$ to be identified with the photon temperature.  The 
thermodynamic quantities technically should be calculated at the temperature
of each particle $T_i$ and then the contributions of each particle to
the density and pressure should be summed.    This
creates a subtlety when a stable particle decouples from the photon
thermal bath yet still contributes to the mass density, pressure,
and entropy of the universe.  When a particle decouples its entropy density
is separately conserved from that of the photon bath.  However, at the
decoupling temperature $T_d$ the sum of the two entropies for $T<T_d$ must
be equal to the total entropy for $T>T_d$.  This is just entropy conservation.
Mathematically this is expressed as 
$s(T_A)T^3_A=s(T_B)T_B^3$
where $T_B=T_d+\delta$ before decoupling of particle $\chi$ 
and $T=T_\gamma=T_\chi$,
and $T_A=T_d-\delta$ is the temperature after decoupling and 
$T=T_\gamma\neq T_\chi$.
However when $\delta \to 0^+$ we can identify $T_A=T$ and $T_B=T_\chi$ 
to obtain
\beq
\label{decoup}
T_i=T\left[ \frac{s(T_d-\delta)}{s(T_d+\delta)}\right]^{1/3}.
\eeq
For stable particles which have decoupled then $T_i$ should be
substituted into the formulas for $\rho(T)$ and $s(T)$.

When additional particles ``distribution decouple'' from the thermal bath
($W$, $h$, etc.\ whose density goes to zero as $T\ll m$) the $T$ in
Eq.~\ref{decoup} is no longer the current photon temperature but rather
the photon temperature before the ``distribution decoupling''.  Using 
conservation of entropy again at the decoupling boundary, we can find
the relation between $T_\chi$ and the current photon 
temperature ($T=T_\gamma$):
\bea
T_\chi & = & \lim_{\delta \to 0^+} T \left[ \frac{s(T_d-\delta)}{s(T_d+\delta)}
   \right]^{1/3}\prod_{T_\gamma < m_i<T_d}
   \left[ \frac{s(T_d^i-\delta)}{s(T^i_d+\delta)}\right]^{1/3} \\
  & =  & T \left[ \frac{s(T_\gamma+\delta)}{s(T_d+\delta)}\right]^{1/3}.
\eea

The thermodynamic quantity 
integrals 
are straightforward to calculate numerically for
any mass and any temperature.  They also can be solved analytically
in the limits that $T\gg m$ and $T\ll m$.  For example,
if $T\gg m$ then the boson energy density is $gT^4\pi^2/30$, and
if $T\ll m$ then energy density decreases rapidly
as $\rho\sim T^{3/2}e^{-m/T}$.  Doing the same for
fermions we can construct the following approximation for $\rho(T)$ 
and $s(T)$:
\bea
\rho(T) & = & \frac{\pi^2}{30}\bar \rho(T)T^4  \\
s(T) & = & \frac{2\pi^2}{45}\bar s(T) T^3 
\eea
where
\bea
\bar \rho(T) & \simeq & \sum_{i=bosons} g_i\theta (T_i-m_i)
            \left( \frac{T_i}{T}\right)^4
           +\frac{7}{8}\sum_{i=fermions} g_i\theta (T_i-m_i) 
                  \left( \frac{T_i}{T}\right)^4 \\
\bar s(T)& \simeq & \sum_{i=bosons} g_i\theta (T_i-m_i)
                   \left( \frac{T_i}{T}\right)^3
           +\frac{7}{8}\sum_{i=fermions} g_i\theta (T_i-m_i)
                   \left( \frac{T_i}{T}\right)^3 .
\eea
Again, for most particles $T_i=T$; stable particles decoupled
from the photon will have $T_i\neq T$.
The equilibrium number density is $n_{eq}(T)\simeq 1.2gT^3/\pi^2$,
$0.9gT^3/\pi^2$ (bosons, fermions)
for $T\gg m$ and decreases rapidly for
$T\ll m$.

\section{Approximating the relic abundance}

Often it is useful to employ approximation techniques to calculate
efficiently and reliably the relic 
abundance~\cite{lee77:165,olive81:497,ellis84:453,kolb90:1}.  
It is based primarily
on dividing thermal history into two distinct eras: before freeze-out,
and after freeze-out.  Freeze-out means that the annihilation rate
of the particle, $\Gamma =n_{eq} \langle \sigma v\rangle$, is less
than the Hubble expansion rate.  Once this occurs the particle can
no longer remain in equilibrium.

From previous sections we have the tools to solve for this freeze-out
temperature given the condition that $\Gamma = H$.  
Massive weakly interacting particles are non-relativistic when they
freeze-out, and so the equilibrium number density can be well
approximated as
\beq
n_{eq} = g \left( \frac{mT}{2\pi}\right)^{3/2} e^{-m/T},
\eeq
where $g$ is the number of degrees of freedom of the relic particle
(2 for a neutralino).
It is convenient to re\"express all temperatures as the dimensionless
variable $x\equiv T/m_\chi$ such that 
\beq
n_{eq} = \frac{g m^3 }{(2\pi)^{3/2}} x^{3/2} e^{-1/x}.
\eeq
Furthermore, from our discussion earlier we know that the Hubble constant
is
\bea
H & = & \frac{\dot a}{a}=\sqrt{\frac{8\pi \rho_\chi G}{3}}
 = \sqrt{N_F} T^2 A  \\
 & = & \sqrt{N_F}m^2 x^2 A
\eea
where $A=\sqrt{8\pi^3G/45}$.  The freeze-out condition 
that $n_{eq}\sigv = H$ at $x=x_F$ yields a transcendental
equation for $x_F$,
\beq
\label{xfeq}
x^{-1}_F =\ln \left[ \frac{g m\sigv (x_F)}{\sqrt{N_F}\sqrt{x_F} 
       A (2\pi)^{3/2}} \right].
\eeq
Eq.~\ref{xfeq} is only logarithmically dependent
on $\sigv$ and yields a value of $x_F\simeq 1/20$ quite consistently
for massive weakly interacting particles.  

After freeze-out the actual number density remains high above the
subsequent would-be equilibrium number density.  It is therefore appropriate
the approximation $n^2-n^2_{eq} \simeq n^2$.
The Boltzmann equation can then be conveniently rewritten as
\beq
\frac{df}{dx}=\frac{m_\chi}{A\sqrt{N_F}}\sigv  f^2
\eeq
where $f=n/T^3$ subject to the boundary condition $f(x_F)= n_{eq}/T_F^3$.
To find the current number density we must integrate this equation from
$x_F$ (freeze-out) down to $x\simeq 0$ (today):
\beq
f(0)=\frac{A\sqrt{N_F}}{m_\chi J(x_F)+A\sqrt{N_F}f^{-1}(x_F)}
\simeq \frac{A\sqrt{N_F}}{m_\chi J(x_F)}
\eeq
where 
\beq
J(x_F)=\int_0^{x_F} dx \sigv (x) .
\eeq

The calculated relic abundance is then simply
\beq
\rho_\chi = m_\chi T^3_\chi f(0) =\frac{A\sqrt{N_F}}{J(x_F)} 
\left( \frac{T_\chi}{T_\gamma} \right)^3 T_\gamma^3 .
\eeq
The ratio $T^3_\chi/T^3_\gamma<1$ is the photon reheating effect from 
entropy conservation, which was calculated earlier.
Scaling this to the critical density $\rho_c = 3H_0^2/8\pi G$ yields
the relic's fraction of critical density.  It is customary to
define the measured value of the Hubble constant
as $H_0=100\, h\, \mbox{km\, s}^{-1}\, \mbox{Mpc}^{-1}$.  Then
\beq
\Omega_\chi h^2 = \frac{A\sqrt{N_F}}{(\rho_c/h^2)J(x_F)}
 \left( \frac{T_\chi}{T_\gamma}\right)^3 T^3_\gamma .
\eeq

We can further reduce the expression for $\Omega_\chi h^2$ by noting
that 
\beq
\left( \frac{T_\chi}{T_\gamma}\right)^3 = \frac{s(T_\gamma)}{s(T_d)}
 \simeq\frac{N_F(T_\gamma)}{N_F(T_d)}=\frac{2}{N_F}.
\eeq
Using this relation and evaluating all numerical constants we get
the convenient form
\beq
\Omega_\chi h^2 = \frac{1}{\mu^2 \sqrt{N_F}J(x_F)}
\eeq
where $\mu=1.2\times 10^5\gev$.  The above formula is an accurate 
approximation to the Boltzmann equation
solution to within about $15\%$.  Table~1 lists the
value of $N_F$ for various ranges of 
the freeze-out temperature $T_F=x_Fm_\chi$. 
\begin{table}
\centering
\begin{tabular}{|cc|}
\hline 
    $T_F=m_\chi x_F$    &  $N_F$ \\
\hline
$m_s -m_c$ & 494/8 \\
$m_c -m_\tau$ & 578/8 \\
$m_\tau -m_b$ & 606/8 \\
$m_b -m_W$ & 690/8 \\
$m_W -m_Z$ & 738/8 \\
$m_Z -m_t$ & 762/8 \\
$>m_t$ & 846/8 \\
\hline
\end{tabular}
\label{tableNF}
\caption{Degrees of freedom corresponding to freeze-out
temperature $T_F=x_F m_\chi \simeq m_\chi/20$.}
\end{table}

Since a massive weakly interacting particle decouples in the non-relativistic
regime, one is able to expand the annihilation cross section
in powers of the relative velocity 
\beq
\sigma v = a+\frac{b}{6}v^2+\cdots 
\eeq
The thermal average~\cite{srednicki88:693} of this expansion yields
\beq
\sigv = a+\left( b-\frac{3}{2}a\right) x + \cdots .
\eeq
With $x_F\simeq 1/20$ this is a quickly converging expansion.  
The constants $a$ and $b$ can be evaluated straight-forwardly with knowledge
of the squared-matrix element of the annihilation 
process~\cite{wells94:219}.  Helicity amplitudes also exist
for all MSSM processes~\cite{drees93:376}.

The power series expansion of the thermal averaged cross-section
breaks down, however, when the relic particle
annihilates into a pole (e.g., a $\chi\chi\to Z$ 
resonance)~\cite{kanekani,griest91:3191}.  One then needs
to use more careful techniques.  Also, the Boltzmann equation becomes
more complicated if there are other particles around with
mass slightly higher than the stable relic (within a ``thermal distance''
$|m_\chi -m_i|\lsim T_F$)~\cite{griest91:3191}.
In this case, the close-by particles can
help restore thermal equilibrium through
coannihilation channels, effectively lowering the relic abundance.
These potentially important cases will not be considered
further here.

\section{Neutralino dark matter}

From the previous sections we have found that $\Omega_\chi h^2$ depends
on $1/\sigv$.  By dimensional analysis we can identify 
$\sigv\sim 1/m^2_{susy}$, which implies that 
$\Omega_\chi h^2\propto m^2_{susy}$.  
Since the age of the universe constraints
are  compatible only with $\Omega_\chi h^2 <1$ (very conservative
requirement) then it should not surprise us that relic abundance
considerations put a upper limit to how large $m_{susy}$ can be.
We shall see quantitatively the results of these constraints in
the following paragraphs.  Before going straight to that,
a few general comments about the lightest neutralino are useful
to review.

The neutralino is a majorana particle, meaning that the particle
and conjugate are the same.  Therefore when two of
them come together to annihilate, Fermi statistics requires each
to be in a different helicity state~\cite{goldberg83:1419}.  
Thus $\chi\chi\to f\bar f$
requires a final state helicity flip for the external fermions,
and the total annihilation rate in the $s$-wave is suppressed
by $m^2_f/m^2_W$ compared to processes that do not require helicity flips.
This is often referred to as ``p-wave suppression'':  the
p-wave configuration doesn't have this Fermi statistics requirement,
although it is suppressed by powers of $v^2$.  
However, at higher values of 
$m_\chi$  other channels start opening up, such as
$W^+W^-$ and $t\bar t$ which do not have this suppression (although
they may still have other coupling constant suppressions).

The majorana nature of the neutralino is one important property
of supersymmetric dark matter.  The other important realization
is that numerous other supersymmetric particles play a role in
the neutralino annihilation channels.  The lightest neutralino, being
the LSP, only annihilates into standard model particles when the
temperature is near freeze-out. However, many important diagrams
of these annihilation processes depend on intermediate supersymmetric
states.  For example, $\chi\chi\to e^+e^-$ depends not only on
an intermediate $s$-channel $Z$ boson, but also on $t$-channel
slepton exchange.  Similarly, $\chi\chi \to q\bar q$ depends on
$t$-channel squark exchange, and $\chi\chi \to W^-W^+$ can depend
crucially on the chargino spectrum.  In general, the neutralino
annihilation rate and therefore the LSP relic abundance depends
on the entire supersymmetric low energy spectrum -- this includes
particle content, masses, and mixing angles.

Upon closer inspection most choices of the supersymmetric spectrum
ultimately give rise to an LSP relic abundance dependent on only
a few parameters in the theory.  
Relic abundance in the minimal model
depends most sensitively on only two parameters, the bino
mass and the right-handed slepton mass.  
As discussed in Martin's introductory article in this volume,
this minimal model is described partly by
a common scalar mass $m_0$ at
the high scale and a common gaugino mass $m_{1/2}$.  As long as
$m_0$ is not much greater than $m_{1/2}$ one typically finds in
the low
energy spectrum a bino (superpartner of the hypercharge gauge boson)
as the lightest supersymmetric particle.  One also typically finds 
the squarks significantly heavier than the sleptons, with
the right-handed sleptons being lighter than the left-handed sleptons.

Taking our hints from the minimal model, 
a useful initial exercise is to declare the LSP as a {\it pure} bino.
Then, the largest
contribution to the neutralino annihilation rate will be from
$t$-channel $\tilde l_R$ exchange in $\chi\chi\to l^-l^+$.
This case has been studied in detail~\cite{drees93:376}, where
it was shown that
\beq
\Omega_\chi h^2 =\frac{\Sigma^2}{M^2 m^2_\chi}\left[
 \left( 1-\frac{m^2_\chi}{\Sigma}\right)^2
  +\frac{m^4_\chi}{\Sigma^2}\right]^{-1}
\eeq
where $M\simeq 1\tev$ and $\Sigma =m^2_\chi +m^2_{\tilde l_R}$.
A good approximation to the requirement that $\Omega_\chi h^2< 1$
yields
\beq
\frac{m^2_{\tilde l_R}}{m_\chi}\lsim 200\gev.
\eeq
This result has two important consequences.  One, it satisfies
the general observation that the relic abundance constraint
puts an upper limit on at least one superpartner.  And second,
in this important example of pure bino, both the bino and the
right-handed lepton superpartner must have masses below
about $200\gev$.  {\it A priori} the upper bound could have been 
any number.
The result is compatible with weak scale supersymmetry
($m_{susy}\sim m_W$), long thought
to be important in electroweak symmetry breaking.

It is not expected that the lightest supersymmetric particle is
a pure bino, but rather mostly bino.  The coupling of the neutralino
to the $Z$ boson depends on the higgsino components.  If the
LSP is partly Higgsino then efficient annihilations through 
the $Z$ boson could be possible, thereby reducing the relic abundance
for a particular $m_\chi$ value.  Limits on the superpartner spectrum
still exist; they just happen to be at different scales  than the
$200\gev$ we found for the pure bino case.

\begin{figure}
\psfig{figure=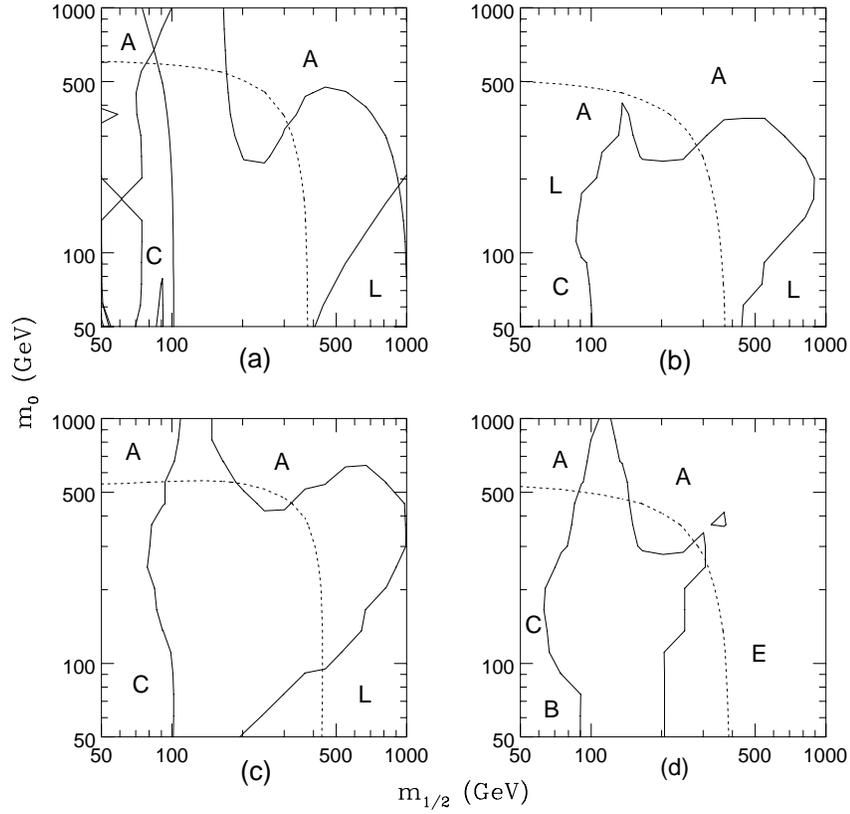,height=4.5in}
\caption{Contours of
allowed parameter space.  $m_t=170\gev$ for each frame.
In frame (a) $\tan\beta =5$, $\sgn (\mu)= -$,
and $A_0/m_0=0$, (b) $\tan\beta=5$, $\sgn (\mu)= -$ 
and $A_0/m_0= -2$, in (c)
$\tan\beta =20$, $\sgn (\mu )=-$ and $A_0/m_0=3$, and in (d) $\tan\beta =10$,
$\sgn (\mu )= +$, and $A_0/m_0 = -2$.  See the text for explanation
of letter labels.
\label{rfig1.ps}}
\end{figure}

Our example realistic particle spectrum is a precise formulation
of the minimal model with electroweak symmetry breaking enforced.
The full $4\times 4$ neutralino mass matrix is diagonalized
numerically, and the relic abundance is calculated.  The four contour
plots~\cite{kane94:6173}
of Fig.~\ref{rfig1.ps} demonstrate the effects of the relic abundance
on the parameter space in the $m_0$ vs.~$m_{1/2}$ plane.
The value of the top quark mass for this contour plot is
$170\gev$ and the sign of the supersymmetric $\mu$ parameter is
chosen to be negative.  In (a) $\tan\beta =5$,
and $A_0/m_0=0$, (b) $\tan\beta=5$ and $A_0/m_0= -2$, in (c)
$\tan\beta =20$ and $A_0/m_0=3$, and in (d) $\tan\beta =10$,
$A_0/m_0 = -2$, and the sign of $\mu$ has been changed to
positive.  The region inside the solid curve is allowed
by all constraints.  Letter labels are placed outside the solid
curve to demonstrate which constraint makes the region 
phenomenologically unacceptable: {\tt A} indicates $\Omega_\chi h^2>1$;
{\tt B} $B(B\to s\gamma)>5.4\times 10^{-4}$; {\tt C}
$m_{\chi^+_1}>47\gev$; {\tt E} electroweak symmetry breaking
problems (full one loop effective potential becomes unbounded
from below); and, {\tt L} the LSP becomes charged
($m_{\tilde l_R}<m_\chi$).  Current limits on $m_{\chi^+_1}$
and $B(b\to s\gamma)$ are somewhat more restrictive than
the limits quoted above, however they will only shrink the
allowed parameter space a little bit near {\tt B} and {\tt C}.
Inside the dotted line corresponds to one particular definition
of acceptable finetuning
allowed in the electroweak symmetry breaking solutions~\cite{kane94:6173}.  
It is not important for our
further discussion.

The peak or spike in the allowed region for $m_{1/2}\simeq 130\gev$
corresponds to $\chi\chi$ annihilations through a $Z$ pole.
Since the LSP is not just a pure bino, and has
some higgsino component to it, the $Z$ pole annihilation
rate becomes extremely important in this region. Also, it
appears obvious from the figures that the relic abundance constraint
has the most effect in the increasing $m_0$ direction.  This is
understandable for several reasons.  Away from the $Z$ and Higgs
poles, the LSP annihilations occur most efficiently through the
$t$-channel scalar diagrams.  For a fixed $m_\chi$ value, which
is roughly
equivalent to fixed $m_{1/2}$, the annihilation cross-section
decreases rapidly as the scalar mass $m_0$ increases.  Hence,
a cutoff in the allowed region must occur as $m_0$ increases.

Forays into the large $m_{1/2}$ region with $m_0$ fixed 
usually end up with other
problems, the most common of which is
$\tilde l_R$ becoming the LSP.  This is simply
a result of the renormalization group equation dependences
on the $U(1)_Y$ couplings.  A quick inspection of the
renormalization group equations for $m_{\tilde B}$ and
for $m_{\tilde l_R}$ proves that if $m_{1/2}\gg m_0,m_W$, then
the right-handed slepton must be lighter than the bino.  
Ruling out this region of parameter space can be considered a relic abundance
constraint, since the dark matter probably cannot 
charged~\cite{basdevant90:395}.

Realization that a weak-scale higgsino could be a legitimate dark matter
particle is a rather recent development.  
One way to obtain an higgsino as the lightest neutralino is
to make $|\mu |$ much less than the gaugino parameters in
the neutralino mass matrix.  A very low value of $\mu$ will 
create a roughly degenerate triplet of higgsinos.  The charged
higgsino and the neutral higgsinos can all coannihilate together
with full $SU(2)$ strength, allowing the LSP to stay in thermal
contact with the photons more effectively, thereby lowering the
relic abundance of the higgsino LSP to an insignificant level.
These coannihilation channels are often cited as the reason why
higgsinos are not viable dark matter candidates.
This claim is true in general, but there are two specific cases
that I would like to summarize below that allow the higgsino
to be a good dark matter candidate.

Drees {\it et al.\/} have pointed out that potentially large 
one-loop splittings among
the higgsinos can render the coannihilations less relevant~\cite{r23}.
Under some conditions with light top squark masses,
one-loop corrections to the neutralino mass matrix will split
the otherwise degenerate higgsinos.  If the mass difference can
be more than about 5\% of the LSP mass, then the LSP will
decouple from the photons alone and not with its other higgsino
partners, thereby increasing its relic abundance.  

Another possibility~\cite{kane96:4458} relating to a higgsino LSP is to 
set equal the bino and wino mass to approximately $m_Z$.  Then
set the $-\mu$ term to less than $m_W$. This non-universality
among the gauginos and particular choice for the higgsino
mass parameter, produces a light higgsino with mass
approximately equal to $\mu$, a photino with mass at about
$m_Z$, and the rest of the neutralinos and both charginos with mass
above $m_W$.  There are no coannihilation channels to worry about
with this higgsino dark matter candidate since no other
chargino or neutralino mass is near it.  
The value of $\tan\beta$ is also required to be
near one so that the lightest neutralino is an almost pure
symmetric combination of $\tilde H_u$ and $\tilde H_d$ higgsino
states.  The exactly symmetric combination does not couple to
$Z$ boson (at tree level).  
The annihilation cross section near $\tan\beta \sim 1$
is proportional to $\cos^2 2\beta$.  The relic abundance scales
inversely proportional to this, and so the nearly
symmetric higgsino in this case is a very good dark matter candidate.
Note that there are no $t$-channel slepton or squark diagrams since
higgsinos couple to sfermions proportional to the fermion mass.  Because
the higgsino mass is below $m_W$, the
top quark final state is kinematically inaccessible, and so the
large top Yukawa cannot play a direct role in the higgsino annihilations.

This non-minimal higgsino dark matter candidate described in the previous
paragraph was motivated by the $e^+e^-\gamma\gamma$ event 
reported by the CDF collaboration at Fermilab~\cite{r26}.  The non-minimal 
parameters~\cite{r26.5}
which leads to a radiative decay of the second lightest neutralino
(photino) into the lightest neutralino (symmetric higgsino) and photon
also miraculously yield a model with a good higgsino dark matter
candidate.

\section{Conclusion}

The minimal model bino and the higgsino described above work as dark
matter candidates both qualitatively and quantitatively.
Nature, of course, might not conform to either of these specific
possibilities, but it is straight-forward to catalog the 
non-minimal possibilities.
There are, of course, numerous ways to go beyond what is presented
here~\cite{gauge}. Scalar mass universality could
be relaxed~\cite{nath97:301}. One could in fact suppose that $R$-parity is not
exactly conserved and the LSP is not a stable particle.  
Perhaps one could 
allow small $R$-parity violations which create meta-stable LSPs whose
lifetimes are greater than the age of the universe to solve
the dark matter dilemma.  However, it is not enough to just make the
lifetime greater than the age of the universe.  Remarkably, the 
measurements of positrons and photons in cosmic rays require
that the LSP lifetimes into these particles
be many orders of magnitude beyond the age of the universe~\cite{ellis92:399}.
In other
words, it is not easy to make
the LSP a meta-stable dark matter candidate.  If $R$-parity were broken
in nature, it appears necessary to look elsewhere for the dark matter
candidate. 

One common theme exists in non-minimal models which 
accommodate a supersymmetric
solution to the dark matter problem.  
It is the requirement that $\chi\chi$ not couple to the
$Z$ boson.  A full strength $SU(2)$ coupling to the $Z$ boson
generally allows too efficient LSP annihilation, and therefore too small
relic abundance to be relevant to the dark matter problem. 
This theme can be found in several non-minimal examples such
as the symmetric
Higgsino~\cite{kane96:4458,r23} described above, 
the $\tilde Z$~\cite{gabutti96:1},
and the sterile neutralino~\cite{carlos97:315} dark matter candidates.
Of course, the minimal model also 
provides a low-strength coupling to the $Z$ naturally
with the mostly bino dark matter 
candidate~\cite{leszek}.  
The older supersymmetry dark matter literature considered
the photino, which also does not couple to the $Z$, as the primary
dark matter candidate~\cite{goldberg83:1419}.

Theoretical
ideas about the
low-energy superpartner spectrum {\it always} have dark matter implications.
The supersymmetric solution to the dark matter problem is perhaps only
superseded by gauge coupling unification in experimentally based
indications that nature might be described by softly broken
supersymmetry.  It is mainly for this reason that much experimental effort
is expended on the search for supersymmetric
relics~\cite{jungman96:195}.  
Likewise, theoretical efforts on understanding
the origin of LSP stability
and the prediction of dark matter properties should
continue to enlighten us.


\section*{References}

\begin{thebibliography}{99}

\bibitem{rparity}
G.~Farrar, P.~Fayet, \PLB{76}{78}{575}; 
S.~Dimopoulos, H.~Georgi, \NPB{193}{81}{150};
N.~Sakai, T.~Yanagida, \NPB{197}{82}{83};
S.~Weinberg, \PRD{26}{82}{287}.

\bibitem{falk94:248}
T.~Falk, K.~Olive, M.~Srednicki, \PLB{339}{94}{248}.

\bibitem{diehl95:4223}
E.~Diehl, G.L.~Kane, C.~Kolda, J.D.~Wells, \PRD{52}{95}{4223}.

\bibitem{trimble87:425}
V.\ Trimble, Ann.\ Rev.\ Astron.\ Astrophys.\ {\bf 25}, 425 (1987);
P.~Sikivie, Nucl.\ Phys.\ Proc.\ Suppl. {\bf 43}, 90 (1995).

\bibitem{schramm}
D.~Schramm, M.~Turner, astro-ph/9706069.

\bibitem{kolb90:1}
R.~Wald, {\it General Relativity}, Chicago: University of Chicago Press (1984);
E.~Kolb, M.~Turner, {\it The Early Universe}, 
Redwood City, USA: Addison-Wesley (1990).

\bibitem{lee77:165}
B.~Lee, S.~Weinberg, \PRL{39}{77}{165};
M.~Vysotskii, A.~Dolgov, Ya.~Zeldovich, Pisma Zh.Eksp.Teor.Fiz.
{\bf 26}, 200 (1977);
P.~Hut, \PLB{69}{77}{85}.

\bibitem{olive81:497}
K.~Olive, D.~Schramm, G.~Steigman, \NPB{180}{81}{497}.

\bibitem{ellis84:453}
J.~Ellis, J.~Hagelin, D.~Nanopoulos, M.~Srednicki, \NPB{238}{84}{453}.

\bibitem{srednicki88:693}
M.~Srednicki, R.~Watkins, K.~Olive, \NPB{310}{88}{693};
P.~Gondolo, G.~Gelmini, \NPB{360}{91}{145}.

\bibitem{wells94:219}
J.D.~Wells, hep-ph/9404219; L.~Roszkowski, \PRD{50}{94}{4842}.

\bibitem{drees93:376}
M.~Drees, M.~Nojiri, \PRD{47}{93}{376}.

\bibitem{kanekani}
G.~Kane, I.~Kani, \NPB{277}{86}{525}.

\bibitem{griest91:3191}
K.~Griest, D.~Seckel, \PRD{43}{91}{3191}.

\bibitem{goldberg83:1419}
H.~Goldberg, \PRL{50}{83}{1419}.

\bibitem{kane94:6173}
G.L.~Kane, C.~Kolda, L.~Roszkowski, J.D.~Wells,
\PRD{49}{94}{6173}.

\bibitem{basdevant90:395}
P.~Smith, in {\it Proceedings of the First International Symposium
on Sources of Dark Matter in the Universe}, edited by D.~Cline,
World Scientific, 1995.
J.~Basdevant, R.~Mochkovitch, J.~Rich, M.~Spiro, A.~Vidal-Madjar,
\PLB{234}{90}{395}.

\bibitem{r23}
M.~Drees, M.~Nojiri, D.~Roy, Y.~Yamada, \PRD{56}{97}{276}.

\bibitem{kane96:4458}
G.L.~Kane, J.D.~Wells, \PRL{76}{96}{4458};
K.~Freese, M.~Kamionkowski, \PRD{55}{97}{1771}.

\bibitem{r26}
S.~Park, ``Search for new phenomena at CDF,'' 10th Topical
Workshop on Proton-Antiproton Collider Physics, edited by
Rajendran Raja and John Yoh, AIP Press, 1995.

\bibitem{r26.5}
S.~Ambrosanio, G.~Kane, G.~Kribs, S.~Martin, S.~Mrenna,
\PRL{76}{96}{3498}; \PRD{55}{97}{1372}.

\bibitem{gauge}
I do not discuss models of low-energy supersymmetry breaking
which could have the gravitino as the LSP, or even exotic messenger
particles as the cold dark matter.  See for example,
S.~Dimopoulos, G.~Giudice, A.~Pomarol, \PLB{389}{96}{37}.

\bibitem{nath97:301}
P.~Nath, R.~Arnowitt, hep-ph/9701301.

\bibitem{ellis92:399}
J.~Ellis, G.~Gelmini, J.~Lopez, D.~Nanopoulos, S.~Sakar, \NPB{373}{92}{399};
G.~Kribs, I.~Rothstein, \PRD{55}{97}{4435}.


\bibitem{gabutti96:1}
A.~Gabutti, M.~Olechowski, S.~Cooper, S.~Pokorski, L.~Stodolsky,
Astropart.\ Phys.\ {\bf 6}, 1 (1996).

\bibitem{carlos97:315}
B.~de Carlos, J.R.~Espinosa, hep-ph/9705315.

\bibitem{leszek}
L.~Roszkowski, \PLB{262}{91}{59}.

\bibitem{jungman96:195}
G.~Jungman, M.~Kamionkowski, K.~Griest, Phys.\ Rept.\ {\bf 267},
195 (1996).

\end{thebibliography}
\end{document}